\documentclass{arxiv-ieeecsmag}

\usepackage[colorlinks,urlcolor=blue,linkcolor=blue,citecolor=blue]{hyperref}

\usepackage{upmath}
\usepackage{comment}

\jvol{XX}
\jnum{XX}
\paper{8}
\jmonth{}
\jname{CiSE}
\pubyear{}

\begin{document}


\title{An RSE Group Model: Operational and Organizational Approaches From Princeton University’s Central Research Software Engineering Group}

\author{Ian A. Cosden}
\affil{Princeton University, Princeton, NJ, USA}



\markboth{Submission to CiSE Special Issue on the Future of Research Software Engineers in the US}{Paper title}

\begin{abstract}
The Princeton Research Software Engineering Group has grown rapidly since its inception in late 2016. 
The group, housed in the central Research Computing Department, comprised of professional Research Software Engineers (RSEs), works directly with researchers to create high quality research software to enable new scientific advances.
As the group has matured so has the need for formalizing operational details and procedures.
The RSE group uses an RSE partnership model, where Research Software Engineers work long-term with a designated academic department, institute, center, consortium, or individual principal investigator (PI).
This article describes the operation of the central Princeton RSE group including funding, partner \& project selection, and best practices for defining expectations for a successful partnership with researchers.

\end{abstract}

\maketitle

\chapterinitial{The breadth } 
and sophistication of software development skills required to build and maintain research software projects are increasing at an unprecedented pace. 
Researchers thrust into the role of developing software, with little or no software development experience or training, often employ ad-hoc, potentially detrimental development methods. 
It has become clear in recent years that the level of effort and required skills to keep pace with computer and programming tools is not in the repertoire of the average researcher \cite{merali}. 
When novices develop software or when researchers are more focused on research publications than on producing high quality software, problems can arise that limit its usability, sustainability, reproducibility, and even accuracy \cite{miller, bhandari}. 
Researchers often lack both the time to refine initial software implementations and the incentives to use best practices while focusing on domain-specific advances. 
As a result, it is not uncommon for software tools and research code to become unusable after a project ends or the primary developer leaves.  

This has led to the emergence of the Research Software Engineer (RSE), a role that is relatively new to both Princeton and the broader research community. 
Coined in 2012, the term Research Software Engineer has been broadly used as an inclusive title to describe anyone who understands and cares about good software and research \cite{baxter}.
By combining an intimate knowledge of research with the skills of a professional software engineer, RSEs have the ability to transform traditional research by providing much needed software development expertise. 
An experienced RSE has the tools and knowledge to allow them to work collaboratively with domain researchers in a manner that ensures the quality, performance, reliability, and sustainability of the software.
At Princeton, an RSE views the development of research software as the primary output of their work efforts, which distinguishes RSEs from domain researchers, who view research publications as the primary focus of their work.
However, due to the relative nascency of the RSE profession, many researchers have not had experience working with an RSE, and because there are not many defined RSE roles, most RSEs are hired into the role without having held an RSE position before. 
Thus, there is a need to provide clear guidance for researchers and RSEs to ensure successful collaborations for all parties involved.

Over the last decade, a number of Research Software Engineering groups have emerged at universities across the world, some becoming quite large and fully integrated into the academic research landscape at their university \cite{katz}. 

Formed in late 2016, the centralized Princeton University RSE group, initially intended to work with researchers to port code to new high-performance computing (HPC) architectures, has experienced significant growth and success by developing and expanding an RSE partnership model. 
Over time, the group has evolved to work on additional research projects far beyond just HPC applications, including data science, social sciences, digital humanities, and even experimental software systems.
In this paper, we introduce the mission of the RSE group, summarize the operational and organizational best practices, and discuss some of the inherent advantages and disadvantages of the central RSE partnership group model.

\subsection{The Princeton RSE Group Mission}
The Princeton RSE group’s mission is to work directly with researchers create the most efficient, scalable, and sustainable research codes possible to enable new scientific and scholarly advances.
This is done by working as an integral part of traditional academic research groups, providing leadership in the design and construction of complex and highly customized software systems. 
The group is committed to creating a collaborative environment in which best software engineering practices are valued, and to sharing and applying cross-disciplinary computational techniques to new and emerging areas.

\section{Organizational Details}
\begin{figure*}
\centerline{\includegraphics[width=26pc]{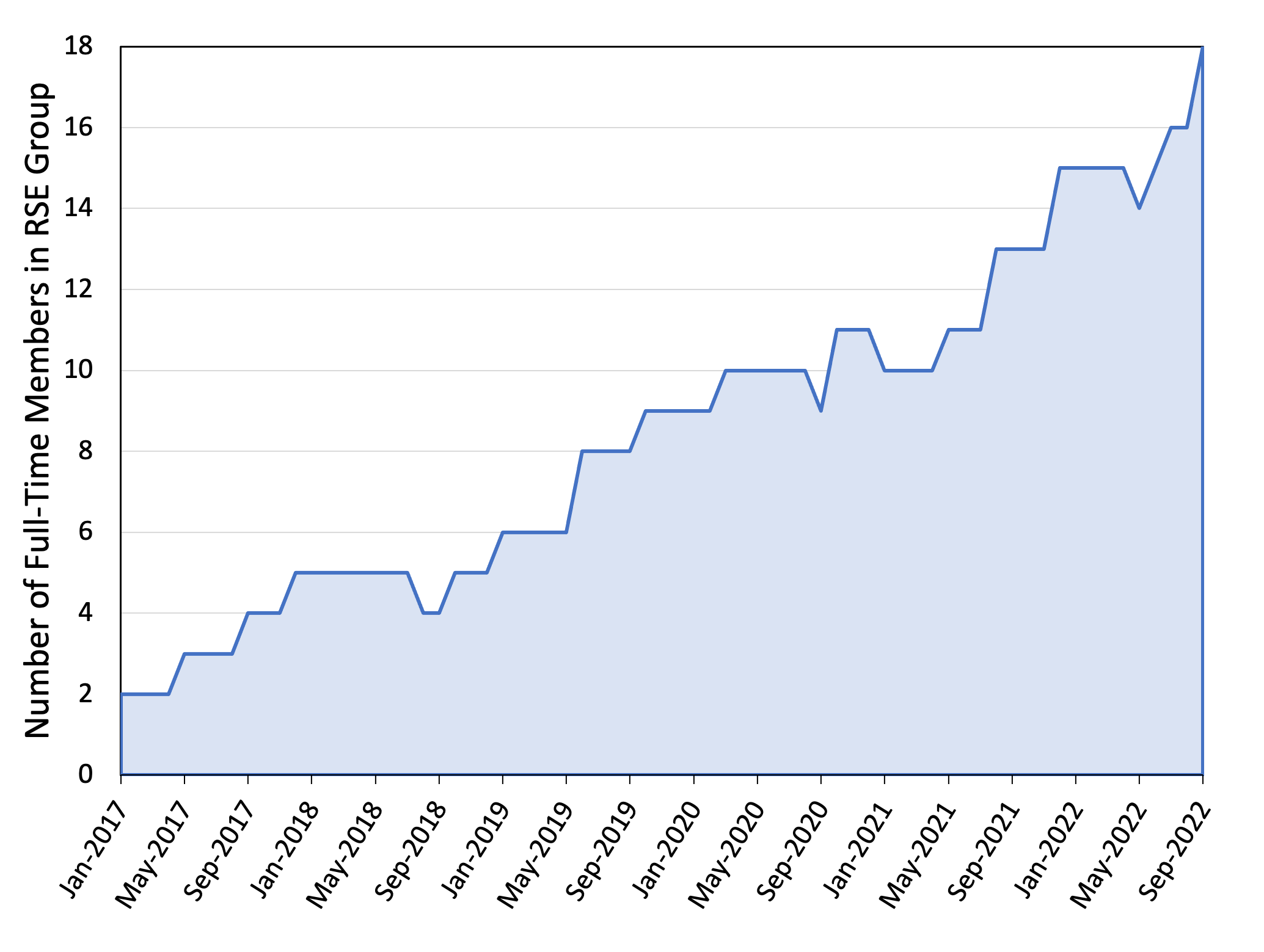}}
\caption{The growth of the Princeton RSE group since January 2017, the month after its inception. This is a count of active members only - not predicted growth or open positions. Down ticks indicate an RSE has left Princeton.}
\label{growth-image}
\end{figure*}

The Princeton Research Software Engineering group is housed with the Research Computing Department within the Office of Information Technology (OIT). 
As of September 2022, the RSE group is led by one Director and two Associate Directors under the Associate CIO for Research Computing. 
The group formed in late December 2016 with two members (a single manager and a single RSE) and has subsequently grown to 18 full-time staff as of September 2022. 
The rate of growth can be seen in Figure~\ref{growth-image}. 
Over the upcoming months, thanks to a broader strategic initiative led by the offices of the Provost and the Dean for Research that aims to invest in selected areas of research, the RSE group will expand by 10 new positions, for a total of 28 members \cite{expansion}. 

Multiple funding mechanisms exist within the group and are described in the section below.
Regardless of funding source, each individual RSE works with a specific research ``partner.'' 
This partner can be an academic department, institute, center, consortium, or individual principal investigator (PI) and are chosen through an internal review process described in the following section. 
An RSE steering committee, comprised of Princeton University faculty members, provides periodic oversight to the group, and serves as the review committee for partner award and selection.


\subsection{RSE Partner Proposal Process and Selection}
As central funding becomes available, a university-wide call for proposals is advertised broadly.
The proposal process is broken into two stages.
First, a one page letter Letter of Intent (LOI) must be submitted explaining how the partner would benefit from increased expertise in research software engineering.
Following an initial screening process, members of the RSE leadership team meet with the submitters to discuss the preparation of a full proposal. 
Depending on the number of received LOIs this is either individuals meetings (when the number of LOI submissions is small) or via town hall information meetings (when the number of LOI submissions is large). 
The purpose of these meetings is to provide guidance to the potential partners on the preparation of the full proposals.
After this initial step - a letter of intent followed by informational meetings - was introduced, the final proposals had a noticeable increase in quality. 
The final proposals must be no more than three pages in length and include: the research area, how a professional RSE can enhance the research, the expertise that would be expected of such a person, the requested appointment level (e.g. RSE vs. Senior RSE), the expected amount of resource needed (e.g., 50\% of an RSE for 3 years), and information about any matching funding that may be provided (e.g., 50\% co-funding).
Co-funding is not required but is encouraged where possible.

\subsection{Proposal Evaluation Criteria}
Proposals are evaluated by the review committee based on the following primary and secondary evaluation criteria.\\

\noindent
Primary:
\begin{itemize}
\item{\emph{Research impact} - Is there a clear opportunity for software-driven impact in the given research area?}
\item{\emph{RSE innovation} - Is there a clear case for which the skills of an RSE are needed to enable the software impact?}
\item{\emph{Deliverables} - Are there clear, achievable objectives for the RSE project for new or improved software?}
\end{itemize}
Secondary:
\begin{itemize}
\item{\emph{Collaborative} - Are there additional collaborative impacts such as  training and engaging students and/or postdocs?}
\item{\emph{Strategic} - Is there an opportunity for impact beyond the specific initial project?}
\item{\emph{Sustainability} - Will there be ongoing impact such as positioning the project and/or PI for external funding, easier subsequent software engagement by students and postdocs, and extensibility and maintainability?}
\end{itemize}

\section{Funding}
Funding for the central RSE program is provided by OIT, the Office of the Provost, and the Office of the Dean for Research. 
The RSE group currently supports two partnership models: Co-Funding and Full Partner Funding.

\textbf{Co-Funding} These RSEs have a home department in Research Computing and report formally to a member of the RSE leadership team (either the Director or an Associate Director). 
A formal relationship is defined between each RSE and their partner. 
These partners agree to fund a fraction (typically 0-50\%) of the position for a finite term (typically 1-3 years) while Research Computing, via one of the offices listed above, provides the remainder. 
If, after the finite term, all parties involved (the partner, RSE, and RSE leadership) agree the partnership can continue indefinitely. 
If, however, the term reaches an end and the partnership ends, the process for finding a new partnership will begin for the current RSE.
It is worth noting that co-funded RSE positions are considered permanent employees of the University and not restricted to the term of the partnership.
This is often an important detail during the recruiting process.

\textbf{Fully-Partner Funded} These RSE positions have a home department outside of Research Computing. 
A partner fully funds an RSE position and commits to the RSE's participation in the RSE group with the same expectations as for other RSE members (including those who are co-funded). 
This includes meeting the expectations listed in the next section. 
The partner and RSE group leadership “co-supervise” the RSE. 
The RSE’s home department and reporting structure will be formally within the partner department/PI and the details of the co-supervision will be agreed upon at the time of hire. 
The fully-partner funded RSEs may have finite term appointments subject to funding availability. 

Regardless of funding method, the expectation is an open and collaborative evaluation between partners to ensure a successful engagement and as such, feedback from partners and all participating PI’s is regularly sought.
 
\section{RSE Expectations}
Regardless of funding, each member of the RSE group is expected to work on the research software projects as prioritized by the partner with an effort of roughly 85\% of their time. 
The RSE group’s leadership team will support the RSE in ensuring that the priorities laid out by the partner are followed.
RSEs are expected to be independent and capable of managing their own time and efforts. 
Awareness of an RSE’s expected weekly effort by the partner and the RSE group leadership will contribute to continued success, and as such, performance evaluations are based on impact and production rather than timekeeping.
In many cases, RSEs are strongly encouraged to track their time. 
This is done on a case-by-case basis depending on the specific situation of each RSE. 
RSEs who are participating in more than one initiative typically track their time to ensure that each project is getting an appropriate effort.

The remaining 15\% of resource time is allocated to RSE group activities and professional development. 
The RSE, along with the RSE group leadership, will track effort to ensure that the 15\% target is met as closely as possible. 
Too much would take away from project work, too little limits growth and minimizes the ability to leverage the collective RSE group knowledge.
These activities include, but are not limited to, those listed below. 

\textbf{Meetings} 
While meetings tend to be minimized, two are essential. 
First, a one-on-one meetings with their direct manager from the RSE group leadership team and a bi-weekly 90-minute RSE group meeting. 
Additionally, a monthly Research Computing all hands meeting occurs once a month.

\textbf{RSE Group Events}
The RSE group organizes occasional internal activities to promote cross-collaboration and sharing. 
Occasional (2-4 times per year) participation in these events is encouraged. 

\emph{RSE project sprints} These short 1-3 day events provide a small team (2-4 RSEs) an opportunity to collaborate on a particularly challenging or technically edifying task. 
This leverages group expertise and provides one project the opportunity to benefit from others as well as an opportunity for RSEs to learn new techniques and strategies. 

\emph{RSE code reviews} The best way to improve code quality is through code review, and one of the best ways to learn new programming approaches is to read other people’s code. 
RSE peers can take turns reviewing each other’s code for specific feedback and advice. 
Both the reviewer and reviewee benefit, as does the project whose code is the subject of the review. 

\textbf{Professional Development}
As largely independent professionals, RSEs must spend time and effort to maintain knowledge on advancing technologies.
Research changes rapidly. 
Software development best practices change even faster. 
Learning is essential in these positions. 
Sometimes this is facilitated by participation in RSE group conversations.
However, a number of other options should be explored on a regular basis including: attending training, tutorials, conferences, contributing to open source projects and other project-based learning opportunities, and collaborating on other Princeton RSE group member projects.

\textbf{Deliver Training}.
As mentioned in the RSE overview, RSEs have knowledge of software engineering best practices that are extremely useful to share with researchers. 
One of the most impactful approaches to sharing such information is through tutorials and workshops. 
These are organized on an individual basis where interested RSEs are encouraged to deliver workshops and mini-courses to share their knowledge with the broader community. 
This is typically considered optional for RSEs, and usually considered only after an RSE has been in their position for more than one year, however, many, if not most, Princeton RSEs participate.

\section{RSE Project Intake, Assignment \& Management}
Allocating individual RSE time to specific projects can be managed in a manner best suited for all parties involved. 
Requirements tend to be unique and therefore the structure of project intake and assignment should be organized in a way that addresses the needs of a partner organization while maximizing the ability of an individual RSE to make a meaningful impact.
An established process for RSE project intake is important, especially when multiple parties are involved.
RSE project intake typically falls into one of the following models:

\subsection{PI project proposal \& review}
For co-funded positions, this is the most typical form of project intake and assignment, especially when the partner is a department or center and has many potential constituents. 
At regular intervals, an email request for software project proposals is sent to associated faculty in the partner department/center. 
Interested faculty respond with a short description of a potential project, and a meeting with the RSE and RSE group leadership is scheduled.
During the meeting, goals and deliverables of the project, an estimated scope, and the expected time commitment are discussed while guidelines for productive collaboration are established. 
After spending time reviewing each proposal, the partner leadership (department chair/manager, etc.) meets with the RSE and RSE group leadership to select successful proposals and define the project commitment. 
The goal is to balance the strategic priorities of the partner and the feasibility of the project with the interests/skills of the individual RSE. 
This commitment is typically for a fraction of the RSE’s availability within a fixed project cycle (typically 6-9 months). 
Regular project check-ins are scheduled throughout the project cycle to ensure project goals are being met. 
Renewals and small maintenance projects are possible through subsequent calls for software projects cycles. 

\subsection{Single project partnership}
For fully-funded partnerships, this is the most common and simplest form of project assignment. 
As the RSE is funded in alignment with a single project, that project is the sole focus of the RSE’s effort. 
As such, there is no need for a process to request, review and select new projects. 
Occasional project check-ins and regular communication are still important to ensure project goals are being met. 

\subsection{Ad-hoc projects} 
Occasionally, an RSE will take on a project by direct request or through a meeting. 
Together with the RSE group leadership and/or the partner representative, they manage the commitment and timeline. 
While this model advantageously allows for flexibility and autonomy of the RSE, maintaining this model increases difficulty with predicting workloads which challenge the ability to advertise to potential PIs and can lead to an overwhelming amount of “walk-ins” and requests.

\section{Defining Expectations}
We strongly encourage communication between partners and both the individual RSEs and the RSE group leadership in the case of questions about the pace of development or the prioritization of effort. 
Several software best practices initially slow down the addition of “new science” but facilitate the development of new aspects of a project while ensuring correctness, reproducibility, and the long-term success of the project.
To retain excellent RSEs, we want to ensure our work environment is as positive and rewarding as possible. 
Enabling intellectual challenges, providing work-life balance, and promoting valued collaboration are all aspects of working at Princeton that keep us competitive. 
If expectations are not being met, then it behooves everyone to quickly address the problem to collaboratively work towards a solution. 
Typically, a new RSE collaboration represents the first time working with an RSE for partners, PIs, and researchers.
As a result, it's proven effective to define the role and expectations ahead of time with all those involved.

A formal \emph{RSE Partnership Guide} is published online and shared with future RSE partners and long-term collaborators.
The purpose of the document is to detail the role of RSEs along with operational details while setting expectations of all those involved. 
It outlines much of what has been discussed here, such as project intake and management and publishing guidelines, as well as fine grained details such as how Annual Performance Reviews are conducted and physical desk locations.
The guide is shared with all potential partners prior to establishing a formal partnership. 
The introduction of the guide has significantly reduced the number of questions and provided a smoother partnership initiation. 

\subsection{RSE Publishing Guidelines}
RSEs sit at the intersection of scholarly research and software development. 
A common and extremely important output of scholarly research is the journal/conference publication. 
RSEs, by design, are not directly measured or graded based on their publication output, however many RSEs, especially those from a pure research background, still value such publications.
We encourage RSEs, PIs, and/or partner managers to discuss authorship as soon as possible. 
It is typically recommended that if a publication would not have been possible without the RSE, then the RSE should be included as a co-author.

RSEs will occasionally be interested in publishing their work in a manner appropriate for the field and individual and RSEs may want to publish their work for a number of reasons. 
In general, we view this as positive and useful for both the individual and the broader community. 
To ensure all parties of an RSE collaboration understand the RSE publishing process, we have published on our website an \emph{RSE Publishing Guidelines} document.
This document gives detailed guidance for publications in which the RSE is either (a) the lead author with or without other members of the research group or (b) a co-author without any other members of the research group. 
The key element of the guideline is frequent, open, and honest communication between the RSE and all parties, including PIs, graduate students, post doctoral researchers, and any other contributors to the research endeavor from which the publication may have resulted. 
The full guidelines are available as part of the previously mentioned \emph{RSE Partnership Guide}.

\section{Advantages \& Disadvantages}
The Princeton RSE model described here developed as a result of Princeton University's unique funding and organizational environment. 
The model has a number of positives, and like any organizational model, some disadvantages. 
They are briefly listed them here for the reader to decide if such a model would be advantageous at a particular institute or organization.
The items listed here are specific to the RSE partner model, rather than more general advantages (e.g. advancing research, improved software sustainability, maintainability, and performance) and disadvantages (e.g. expense, recruiting, retaining) of Research Software Engineers.

\subsection{Advantages} 
\textbf{Embedded Expertise} The partnership model has the distinct advantage that the scope and domain are both known ahead of time and stay relatively fixed over a long period.
The result is a customized job description and hiring search that allows for a very specific set of skills and knowledge in a successful candidate.
Then, over time, RSEs acquire domain expertise that makes them increasingly valuable to the projects and partners. 
We've observed that after 1-2 years with the same project/partner that the depth of contribution possible by RSEs increases significantly. 

\textbf{Group Management} Hiring RSEs and the subsequent management can take a significant amount of effort - something that researchers and partners frequently would not have time to adequately devote to. 
RSEs managed and supervised by a dedicated RSE leadership team that prioritizes the entire RSE endeavor results in more complete hiring searches, increased focus on professional development, and technical support for individual RSEs.
This aspect is one of the initial appeals of the partnership model and a major reason fully-partner funded RSEs join the group.

\textbf{RSE Collaboration} The centralized nature of the RSE group allows for easy sharing, collaboration, and cooperation amongst the group members.
The group activities mentioned above (sprints, code review, etc.) are routinely cited by the group members as a unique opportunity to learn new technologies and develop new skills by working on a real and meaningful project.

\textbf{Sharing Best Practices} A frequent comment from researchers is that over time RSEs begin to serve as leaders and mentors to novice software developers, including undergraduates, graduate students, and postdoctoral researchers. 
By providing mentorship on research software projects, RSEs elevate domain scientists’ development through exposure to professional best practices often neglected in research software (e.g. version control, building unit tests, and documentation). 
Also, RSEs within the group have taken an active role in designing and delivering training courses and workshops on research software engineering best practices. 
Sessions typically range from 1-4 hours and targeting students and researchers, include topics ranging from continuous integration and automation, advanced git, to performance tuning. 
These training programs are uniquely beneficial as RSEs understand the technology, audience, and requirements of research software.

\subsection{Disadvantages}
\textbf{Single Point of Failure} While the long-term commitment to specific partners allows an RSE to develop a deep collaboration and make increasingly meaningful contributions, it also leaves the project vulnerable should the RSE leave the position.
RSEs have marketable skills, desirable in other research settings and industry, which inevitably means that there is turnover withing RSEs.
When an RSE leaves a project it can leave a massive hole in the project as after years the RSEs can become nearly impossible to replace. 
The gap between when an RSE leaves and when a new RSE can be hired and onboarded is time when project work can become completely stalled. 

\textbf{Lack of Flexibility} The partnership model limits the ability to take projects outside the given partnership regardless of fit or need. 
For example, if a specific RSE would be better suited for a project with \emph{Partner B} but works with \emph{Partner A} it's extremely difficult to switch to \emph{Partner B}.

\section{CONCLUSION}
The Princeton Central RSE group's growth is a testament to its impact on advancing and accelerating research.
What started as something of an experiment has grown well beyond the proof-of-concept stage straight into the strategic investment stage.
While the partnership model has proven incredibly effective, as the group has grown, so has the need to address some of the weaknesses of the model discussed earlier.

\section{ACKNOWLEDGMENT}

The author would like to thank all the Princeton community members who have contributed to the success of the program in its early years. 
The list is too long to write out, but includes partners from virtually every academic department, PICSciE, and OIT, each of whom have consistently contributed ideas and effort to help instantiate this new RSE initiative.
I would also like to directly thank each individual RSE member of the group. 
The success of the program is a direct result and testament to their sustained hard work and the transformative impact they have had on project after project.  
Finally, I'd like to gratefully acknowledge and thank Curt Hillegas, Jeroen Tromp, and Jay Dominick for the consistent and overwhelming support of the RSE group at Princeton. 
Without them, the RSE group would not just be less successful, it wouldn't exist.

\begin{IEEEbiography}{Ian A. Cosden}{\,}is the Director for Research Software Engineering for Computational and Data Science at Princeton University, in Princeton, NJ.
He earned a Bachelors in Mechanical Engineering from University of Delaware, an M.S. in Mechanical Engineering from Syracuse University, and a Ph.D. in Mechanical Engineering from the University of Pennsylvania.
At Princeton, he leads a team of Research Software Engineers (RSEs) who complement multiple traditional academic research groups by offering embedded, long-term software development expertise. 
Additionally, he is the current and founding chair of the Steering Committee for the US Research Software Engineer Association (US-RSE) . 
Contact him at icosden@princeton.edu.
\end{IEEEbiography}

\end{document}